The effect of paternal age on offspring intelligence and personality when controlling

for paternal trait level

Ruben C. Arslan[1], Lars Penke[2], Wendy Johnson[3], William G. Iacono[4], & Matt

McGue[4]

[1] Institute of Psychology, Humboldt University of Berlin

[2] Institute of Psychology, Georg August University Göttingen

[3] Department of Psychology and Centre for Cognitive Ageing and Cognitive

Epidemiology, University of Edinburgh

[4] Department of Psychology, University of Minnesota

Corresponding author: Ruben C. Arslan

Solmsstr. 29

10961 Berlin

Germany

Email: ruben.arslan@gmail.com

Telephone: +49 163 8659876

**Abstract**

Paternal age at conception has been found to predict the number of new genetic mutations. We examined the effect of father's age at birth on offspring intelligence, head circumference and personality traits. Using the Minnesota Twin Family Study sample we tested paternal age effects while controlling for parents' trait levels measured with the same precision as offspring's. From evolutionary genetic considerations we predicted a negative effect of paternal age on offspring intelligence, but not on other traits. Controlling for parental IQ had the effect of turning a positive-zero order association negative. We found paternal age effects on offspring IQ and MPQ Absorption, but they were not robustly significant, nor replicable with additional covariates. No other noteworthy effects were found. Parents' intelligence and personality correlated with their ages at twin birth, which may have obscured a small negative effect of advanced paternal age (< 1% of variance explained) on intelligence. We discuss future avenues for studies of paternal age effects and suggest that stronger research designs are needed to rule out confounding factors involving birth order and the Flynn effect.

*Keywords*: paternal age, intelligence, personality traits, evolutionary psychology, mutation load, de novo rare variants, mutation-selection balance

## 1. Introduction

The well-established genetic influences on psychological traits such as intelligence and personality traits have attracted the attention of a growing number of evolutionary psychologists. This is because selection continues to exert pressure on heritable traits, unless they are completely irrelevant for fitness. There is evidence that neither intelligence nor personality traits are currently completely neutral to selection, but are associated with fitness components like survival [1–3] and reproductive outcomes [4–7]. If we make the assumption that at least somewhat similar selection processes affected these traits with some consistency during the last few thousand years we must wonder why genetic differences in these traits persist. They imply the existence of maintaining evolutionary mechanisms, since otherwise natural selection would drive variants to fixation and eliminate the differences [8,9]. Here we tested the hypothesis that harmful genetic mutations that occur anew each generation might contribute to the genetic variation particularly of intelligence, which would suggest that this genetic variation is maintained by a balance of mutation and counteracting selection. To test this hypothesis, we relied on paternal age at twin birth (henceforth simply "paternal age") as a proxy of new mutations and used a better-controlled design than previous studies. We review the increasingly supportive evidence for paternal age as an indicator of new mutations as well as the importance of using the right controls after explaining the evolutionary genetic reasoning behind the hypothesis that mutations contribute substantially to the genetic variation in intelligence.

### 1.1. Evolutionary Explanations for Individual Differences

Because intelligence is regarded as an attractive trait in mates across cultures [10–13], it is plausible that higher intelligence was also preferred during recent human evolutionary history. Thus, high intelligence could be positively sexually selected, driving low intelligence to extinction (barring evolutionarily very novel impediments like effective birth control; [14,15]). There is also evidence for survival selection for intelligence in current times [3], though it is of course hard to infer how the relation between intelligence and survival has varied during evolutionary history. To explain why high intelligence has not been fixated, Penke et al. [9] argued that intelligence has a large number of relevant genetic loci and thus presents a large target for mutations. Mutational target size includes loci that are not polymorphic, but whose alteration would affect the trait. It thus includes a larger number of loci than those which are currently polymorphic and might be picked up by analyses of common single nucleotide polymorphisms (SNPs). As such a target, intelligence would be under *mutation-selection balance*, which occurs when purifying selection removes mutations deleterious to fitness from the gene pool, but cannot outpace the occurrence of new mutations. Thus, a number of mutations persist in the population and individuals have varying "mutation loads" [16]. Depending on whether intelligence is assumed to be under directional or stabilising selection, predictions can be made with regard to the effect of genetic perturbations. Stabilising selection would lead to a buffering against both deleterious and beneficial changes, whereas under directional selection a higher evolvability or responsiveness to perturbations would be predicted [17]. Higher robustness would imply smaller effects of new mutations [18].

Penke et al. [9] contrasted personality traits with intelligence to show that they are not just distinct due to convention or methods, but a product of different selection pressures. They argued that personality variation does not fit the pattern evoked by mutation-selection balance and predicted only a medium mutational target size for personality variation. They suggested that this favoured *balancing selection* as the explanation for personality differences. Balancing selection is a class of mechanisms in which fitness effects of a trait variant differ by environment, be it spatial, social, temporal or genetic (i.e. epistasis and overdominance) [19]. Variation is thus maintained by selection of different trait levels in different environments.

Recent genome-wide complex trait analyses found that more of the genetic variation of intelligence [20–23] than of personality traits [24,25] is associated with small genetic relationships captured by common genetic variants that have high frequencies in the population and are thus unlikely to be novel harmful mutations. Despite this, some of the genetic variation remains unexplained in these traits. Furthermore, the weak signals from the common genetic markers in these studies might come from older mutations in linkage disequilibrium with the markers, or even from novel rare variants in weak linkage disequilibrium but with strong effects on the phenotype [26–28]. In fact, we know of more than 300 of rare mutations that have major effects on intellectual ability [29–31]. Potential participants with intellectual disability are usually excluded from research on intelligence in the normal range. Still, substantial evidence points toward mutation-selection balance [32] as a reasonable explanation for much of the genetic influence on intelligence. However, these and other molecular genetic findings [32–34] cast doubt on balancing selection as the main explanation for genetic personality variance. Instead, a more differentiated view of different personality domains is spreading [24,35]. Lukaszewski and Roney [36] argued that extraversion might be calibrated to attractiveness and physical formidability, which are probably under mutation-selection balance. Similarly, Miller [37] invoked costly signalling theory, with costly signals calibrated to exhibit high fitness, to account not only for genetic variation in intelligence, but also in generosity, courage and even agreeableness and conscientiousness. Still, in the absence of a new convincing pattern implicating a specific mechanism in personality, balancing selection may still be a viable explanation.

## 1.2. Genetic Mutations and Paternal Age

Mutation-selection balance can occur because mutations are generally much more likely to harm the intricate system they affect than to add adaptive benefits to it [38], so the expected effect of new mutations is in the opposite direction of selection. But where do new mutations originate? To maintain mutation-selection balance, mutations need to be inherited, so they need to be germline, not somatic, mutations. Keightley [38] estimated an average of about 70 new germline mutations per human each generation, which is in line with Kong et al. [39], who reported an average of 63.2 new mutations when comparing the sequenced whole genomes of parent-offspring trios. Keightley [38] also estimated that on average 2.2 of these new mutations per generation are deleterious (reducing fitness), which would be implausibly high if each mutation had to be eliminated through failure of the carrier to reproduce, but not if selection acts on relative fitness differences among individuals (quasi-truncation selection [40,41]).

Keightley [38] reviewed the available evidence and found that most mutations are paternal in origin, as had been suggested for a long time [42–44]. His finding was corroborated by Campbell et al. [45] and Kong et al. [39]. The latter reported 3.9 times higher mean single nucleotide mutations of paternal than maternal origin. Strikingly, the number of paternal-origin mutations appeared much more heterogeneous than the number of maternal-origin mutations (ratio of variances = 8.8). The heterogeneity in male mutation rates could almost entirely be accounted for by paternal age; Kong et al. [39] reported an estimated increase in paternal mutations with age of two per year. Crow [46] found single nucleotide mutations in which one nucleotide had been mis-transcribed into one of the other three to be far more common during male than female gametogenesis. Originally, the suspected reason for this was the far greater number of pre-meiotic cell divisions in sperm (35 + 23 * years after puberty) compared to oocytes (24) leading to an accumulation of errors with age. New data is consistent with a linear relationship, but there is also evidence for "selfish spermatogonial selection" (i.e. pre-meiotic selection for mutated cells) at a few loci [47–51].

Decay of transcription fidelity, proofreading error, or some combination of these pathways [46,52] may also be involved.

Single nucleotide mutations appear to be the most common type [53], though they do not account for the most altered base pairs per birth [47]. Unlike chromosomal aberrations such as trisomy 16 and 21, they do not occur more often with advancing maternal age [38,39,54]. Like single nucleotide mutations, new copy number variants (CNV; duplicated or deleted base pair sequences) also seem to have a paternal origin bias [47] and to be associated with increasing paternal age in mouse models [55]. Three recent molecular genetic analyses found no significant relation between new CNVs in offspring and parental ages at birth. Two analysed the exomes (i.e. the coding 1% fraction of the genome) of clinical samples and have limited statistical power and generalizability [56,57], but a third [58] found neither associations of paternal age with rare CNVs, nor with several proxies for new CNVs in large samples of either healthy participants or schizophrenia patients. On the other hand, Hehir-Kwa et al. [59] found a paternal origin and age bias specifically for new CNVs with non-recurrent breakpoints in a large cohort of individuals with intellectual disability.

Because epigenetic insults accumulate in somatic cells during a lifetime [60], there has been speculation that paternal age effects could potentially be explained through epimutations [61,62], though erasure of epigenetic information in the germline is thought to limit if not prevent their inheritance [60]. Though it will be difficult to identify each factor's independent contributions and to disentangle cases where genetic mutations cause epigenetic changes from cases where epigenetic instability causes genetic mutations [63], current evidence [60] points towards genetic mutations as the main explanation for subsequent paternal age effects.

To summarize, since paternal age at conception is linearly related to the number of pre-meiotic cell divisions, it can be used as a proxy for likelihood of new germline mutations [39].

*1.3. Paternal age and psychological traits*

Keller and Miller [64] and Uher [65] argued that severe mental illnesses that confer strong reproductive disadvantages should owe their continued existence to pleiotropic effects of rare recent mutations. Indeed, the increased likelihood of schizophrenia in offspring of older fathers is well documented [66] and has been noted since the 1950s [67] and more recently by Malaspina et al. [68]. Reichenberg et al. [69] reported similar observations for autism, as did Frans et al. [70] for bipolar affective disorder and Lopez-Castroman et al. [71] for intellectual disability. By contrast, effects seem to be trivial or zero for unipolar depression [72]. Paternal age associations with sporadic (nonfamilial) cases of Apert's syndrome, achondroplasia, progeria and other diseases have been found consistently [65]. For autism, schizophrenia and intellectual disability, paternal age effects have recently been corroborated by exome-sequencing studies [31,73–75], some of which also reported auxiliary analyses of the association of paternal age with IQ.

Searching for an endophenotype of schizophrenia led Malaspina et al. [76] to examine the relation of paternal age with IQ in the general population. In a large ($N = 44,175$) sample of Israeli conscripts, they reported a shallow inverted U-shaped relation with IQ (especially non-verbal IQ), which is a risk factor for schizophrenia [77]. The relation persisted even after controlling for maternal age, parental education and numerous other possible confounds. One subsequent, independent study replicated the finding in a large ($N = 33,437$) sample of children for several intelligence measures across three waves [78]. However, after controlling for maternal education, birth order, birth weight and family size in the same sample, Edwards and Roff [79] found many of the associations reduced to non-significance. They argued for the added controls, but Svensson, Abel, Dalman, and Magnusson [80] expressed concern that the correction for birth weight might remove a mediated effect [81,82] and make a real association look spurious. Still, the largest effect reduction resulted from controlling for maternal education. This choice can hardly be contested, because maternal education can be a proxy for heritable maternal intelligence and Edwards and Roff [79] showed that maternal education correlated negatively with father's age due to period effects on education in their cross-sectional sample. Svensson et al. [80] did not find a negative link between paternal age and scholastic achievement in adolescence either. Their sample (the largest so far; $N = 155,875$) comprised recent birth cohorts in Stockholm county, where delayed paternity is common. They controlled maternal and paternal education (not scholastic achievement), country of birth, parental mental health service use and graduation year (to control rising grades).

Auroux et al. [83], who had previously reported a negative association between advanced paternal age and military aptitude test scores [84], did not replicate that relation in newer data ($N = 6,564$) when controlling for parents' academic level, but instead found an association between lower paternal age and lower aptitude. Similarly, Whitley et al. [85] found lower IQs for children born to younger fathers after controlling for number of older siblings ($N = 772$). However, in the same sample they found no association between paternal age and reaction time, arguably a measure less influenced by cultural, social and educational background.

Statistical controls for parental traits are still necessary when new mutations are directly quantified. However, in three recent clinical exome-sequencing studies, such controls were not possible and the reported associations with intelligence may thus have been biased: Iossifov et al., [86] and Sanders et al. (2012) [54] counted new SNPs by comparing parents' and children's exomes. They reported no links between new rare SNPs and intelligence. Sanders, et al. (2011) [56], on the other hand, reported a negative association with CNVs. Generalizability may be limited here because the children had autism spectrum disorders. In an earlier study using SNP arrays [87], rare CNV burden was found to predict intelligence in a small clinical sample. This association was not replicated in two larger, nonclinical samples by Bagshaw et al. [88] (2013) and by McRae, Wright, Hansell, Montgomery, and Martin [89]. Intellectual disability, which is excluded from most studies of IQ in the normal range (but see [90]), has been linked to new CNVs on several occasions [31,59,91]. Rauch et al. [31] estimated new SNPs to explain up to 55% of cases of non-syndromic, sporadic intellectual disability in a small exome-sequencing study.

The same research groups who found negative links between paternal age and intelligence also reported associations in the same general population samples between paternal age and aspects of personality, namely poor social functioning [92] and externalising behaviour [93]. Lundström et al. [94] reported a U-shaped relation of paternal age with autistic-like normal variation in two Swedish twin samples. However, Robinson, Munir, McCormick, Koenen, and Santangelo [95] did not find an effect of parental ages on extreme social-communicative autistic traits in the general population. Kuja-Halkola, Pawitan, D'Onofrio, Långström, and Lichtenstein [96] reported an association between advanced paternal age and number of violent crimes in a sample of more than two million individuals. Importantly, the association persisted when differentially exposed siblings were compared in a within-family analysis, thereby controlling for factors shared by siblings. Plausibly, personality and intelligence might mediate this relationship [97].

An explanation of the paternal age effects that relies on new mutations or epigenetics [62] mandates thorough control for alternative explanations. An obvious possibility is that parental personality [98] and intelligence [15] influence reproductive timing and therefore paternal age. Offspring's inherited personality and intelligence would then differ according to paternal age because of this unobserved common cause. So far, parental intelligence and personality as confounds have not been ruled out, because only proxy variables like education or socioeconomic status (SES) were available in the samples. Proxies for personality measures have not yet been controlled in any study, to our knowledge.

The effect of controlling for family predisposition has been studied more when it comes to mental illnesses. If genetically vulnerable individuals delay reproduction, the fathers of familial cases should be older than those of sporadic cases, whose relatives did not have similar illnesses. The reverse was found for several mental illnesses (see e.g. [99,100]; but also a non-replication by Pulver et al. [101]). Still, advanced paternal age could also exacerbate a pre-existing familial vulnerability to mental illness (i.e., be a factor in familial cases as well) or an existing familial vulnerability could be masked through hastened reproduction by vulnerable individuals. If parental vulnerability to mental illness could be accurately quantified and controlled, effects of new mutations could be isolated. With continuously measured traits like intelligence and personality, we can hope to control for the parental contribution with greater precision.

*1.4. The Present Study*

We addressed several of the limitations of prior studies in a large, population-based twin and family sample. To isolate the effect of new mutations from the expected, inherited trait level, we controlled for parental intelligence or parental personality traits when assessing the influence of paternal age on these traits in the offspring. We also controlled for birth order, which was correlated with paternal age, to account for the possibility of diminishing parental investment in later-born children [102,103]. Though a twin sample was not strictly necessary for our purposes, we used the co-twins to replicate all results found in one half of the sample. Samples with detailed parental and offspring trait measurements are valuable but rare, which  offsets potential problems with generalizability to singletons [104,105].

On theoretical grounds and based on previous results, we predicted a small remaining negative paternal age association with offspring intelligence after applying these controls. We also looked for paternal age associations with offspring head circumference as a proxy for brain size [106–109]. Head circumference is highly heritable [110], but not highly correlated (about .10-.20 [107,111]) with intelligence. While brain size has been discussed as an evolutionary proxy for intelligence [111], there is evidence indicating that, unlike intelligence, it may have been under stabilising selection [109]. Therefore, we did not expect to find a paternal age association with head circumference.

For personality traits, on the other hand, we did not expect to replicate the association between paternal age and offspring externalising behaviour and social functioning reported by Saha et al. [93] and Weiser et al., [92] when using analogue personality traits and controlling for parental personality trait levels. In these studies proxy variables for the parental trait levels were not controlled. Therefore it cannot be ruled out that parental personality affected reproductive timing [98], which could have introduced a spurious association between paternal age and the children's personality. Absence of association after control would be consistent with the theoretical prediction that personality traits are mostly not under mutation-selection balance [9], though it would not provide direct evidence for the absence of mutation-selection balance.

## 2. Methods

### 2.1. Sample

The sample comprised 1,898 pairs of same-sex twins (52% female; 64% monozygotic) and their parents who participated in the intake assessment of the Minnesota Twin Family Study (MTFS), an ongoing population-based longitudinal study. State birth records provided the starting point to locating more than 90% of all Minnesotan same-sex twins born in the target periods. Twins with birth defects and major disabilities were screened out of the sample. Less than 20% of the located families declined participation. Based on a brief survey which 80% of the decliners completed, it was possible to show that decliners were only slightly less educated (<0.3 years) and did not differ from participants with regard to self-reported mental health. At intake two thirds of the assessed twins were approximately 11 years old (born 1977-1994) and one third were approximately 17 years old (born 1972-1979). Like the population of Minnesota in the periods of their births, the twins predominantly (over 95%) had European ancestry. Iacono, Carlson, Taylor, Elkins, and McGue [112] and Iacono and McGue [113] described the recruitment process and the characteristics of the sample in more detail. The 11-year-old cohort was enriched for twins showing antisocial behaviors by recruiting pairs in which at least one showed symptoms of attention-deficit-hyperactivity disorder or conduct disorder [114]. About 11% of participants were recruited in this way; we refer to them as the "enrichment sample". Neither attention-deficit-hyperactivity-disorder [115] nor conduct disorder [116] has been linked to paternal age.

### 2.2. Ethics Statement

The University of Minnesota's institutional review board approved the collection of the data used in this study. Twins gave written informed assent and parents gave written informed consent.

### 2.3. Measures

Twins' birth dates were available from state records. Their zygosity was assessed based on the consensus of several methods and serological analyses in case of disagreement.

In the intake phone survey, the mother reported the father's birth date and education, the twins' birth weight, any birth complications and whether the twin birth had been full-term or by how many weeks it had been early or late. If the father had taken part in the intake assessment, we used his self-reported birthdate and education data instead.

The 11-year-old-cohort of twins was assessed at intake using an abbreviated Wechsler Intelligence Scale for Children – Revised (WISC–R). It comprised two verbal (Vocabulary and Information) and two performance (Block Design and Picture Arrangement) subtests, which had been selected to maximize the correlation (.90) with the full WISC–R. The 17-year-old-cohort and the parents were assessed using the same subtests of the Wechsler Adult Intelligence Scale–Revised (WAIS–R). Altogether, 1,531 families had complete IQ data (see Table 1 for $n$s for each family member as applicable for our analyses).

Both cohorts completed the eleven primary scales of the Multidimensional Personality Questionnaire (MPQ) at approximately age 17 years. Their parents completed the questionnaire at intake ($n$s = 1,109 families with complete data for the superfactors, $n = 1170$ for Absorption). The MPQ primary scales can be aggregated into three superfactors (Positive Affectivity, Negative Affectivity, and Constraint) plus an Absorption factor. Positive Affectivity comprises the scales Well-being, Social Potency, Achievement and Social Closeness. Negative Affectivity contains the scales Aggression, Alienation, and Stress Reaction. Constraint consists of Control, Traditionalism, and Harm Avoidance [117]. A joint factor analysis by Church [118] of Tellegen's personality model with the popular Big Five model revealed no gaps in coverage of either instrument in comparison with the other.

Head circumference ($n = 1,225$ families) was measured during the intake assessment.

Fathers who did not take the intelligence test ($n = 336$) had been educated fewer years (Cohen's $d = -0.26$, $p < .001$). Their twins had significantly lower IQs (-0.17, $p < .001$), less constraint (-0.06, $p = .045$), less positive (-0.07, $p = .019$) and more negative affectivity (0.10, $p < .001$). Mothers in these families also had significantly lower IQs (-0.09, $p = .014$), less constraint (-0.09, $p = .022$) and more negative affectivity (0.09, $p = .016$).

**Table 1:** Descriptive statistics for main variables

| Variable | Individual | $n$ | Mean | $SD$ | Range |
|---|---|---|---|---|---|
| Paternal age | Twins | 1848 | 30.15 | 5.54 | 15-53 |
| Birth weight (gms.) | Twins | 3700 | 2587 | 563 | 566-4961 |
| Nr. of older siblings | Twins | 3704 | 0.94 | 0.93 | 0-8 |
| Nr. of younger siblings | Twins | 3704 | 0.85 | 0.92 | 0-9 |
| Birth year | Father | 1860 | 1952.14 | 7.46 | 1925-1977 |
|  | Mother | 1888 | 1954.42 | 6.90 | 1934-1976 |
|  | Twins | 3759 | 1982.26 | 6.07 | 1972-1994 |
| IQ | Father | 1562 | 106.53 | 14.67 | 61-151 |
|  | Mother | 1851 | 102.29 | 13.44 | 70-147 |
|  | Twins | 3749 | 102.23 | 13.93 | 50-156 |
| Education (years) | Father | 1859 | 14.17 | 2.51 | 6-26 |
|  | Mother | 1879 | 13.97 | 2.06 | 6-24 |
|  | Twins | 3277 | 7.49 | 2.75 | 2-16 |
| Head circumference (cms.) | Father | 1362 | 578.81 | 17.15 | 531-640 |
|  | Mother | 1614 | 554.36 | 18.45 | 455-767 |
|  | Twins | 2927 | 548.45 | 20.81 | 228-613 |
| MPQ Positive affectivity | Father | 1514 | 120.64 | 13.10 | 76-162 |
|  | Mother | 1735 | 120.10 | 13.12 | 65-164 |
|  | Twins | 2931 | 123.17 | 13.42 | 64-166 |
| MPQ Negative affectivity | Father | 1514 | 82.17 | 13.54 | 41-130 |
|  | Mother | 1735 | 80.81 | 12.89 | 38-122 |
|  | Twins | 2931 | 88.91 | 14.27 | 42-147 |
| MPQ Constraint | Father | 1514 | 144.03 | 14.55 | 87-186 |
|  | Mother | 1735 | 151.18 | 13.61 | 100-195 |
|  | Twins | 2931 | 134.09 | 16.14 | 58-187 |
| MPQ Absorption | Father | 1538 | 38.52 | 8.60 | 18-67 |
|  | Mother | 1762 | 41.28 | 9.16 | 18-69 |
|  | Twins | 2976 | 42.70 | 9.42 | 18-72 |

Note. Total N = 1898 families

*2.4. Statistical Analyses*

We fitted structural equation models (SEMs) using Mplus version 7 [119] with a robust maximum likelihood estimator. Compared to standard multiple regressions, full information maximum likelihood (FIML [120]) allowed us to use all available data, not just complete cases, and by using latent variables we were able to estimate comparable regression coefficients, indicating expected change in outcome in standard deviation units per decade of paternal age, across outcomes with different reliabilities.

We fitted two separate but analogous chains of models for intelligence and head circumference in one chain, and MPQ personality traits in the other. In the intelligence models we let the residuals of the verbal subtests Vocabulary and Information and those of the performance subtests Picture Completion and Block Design correlate. We allowed subscale residuals to correlate between the twin-pairs to allow for similarity greater than expected on the basis of the latent factors. We also let the Absorption factor correlate within the superfactors Positive and Negative Affectivity. A simplified model for IQ can be seen in Figure 1.

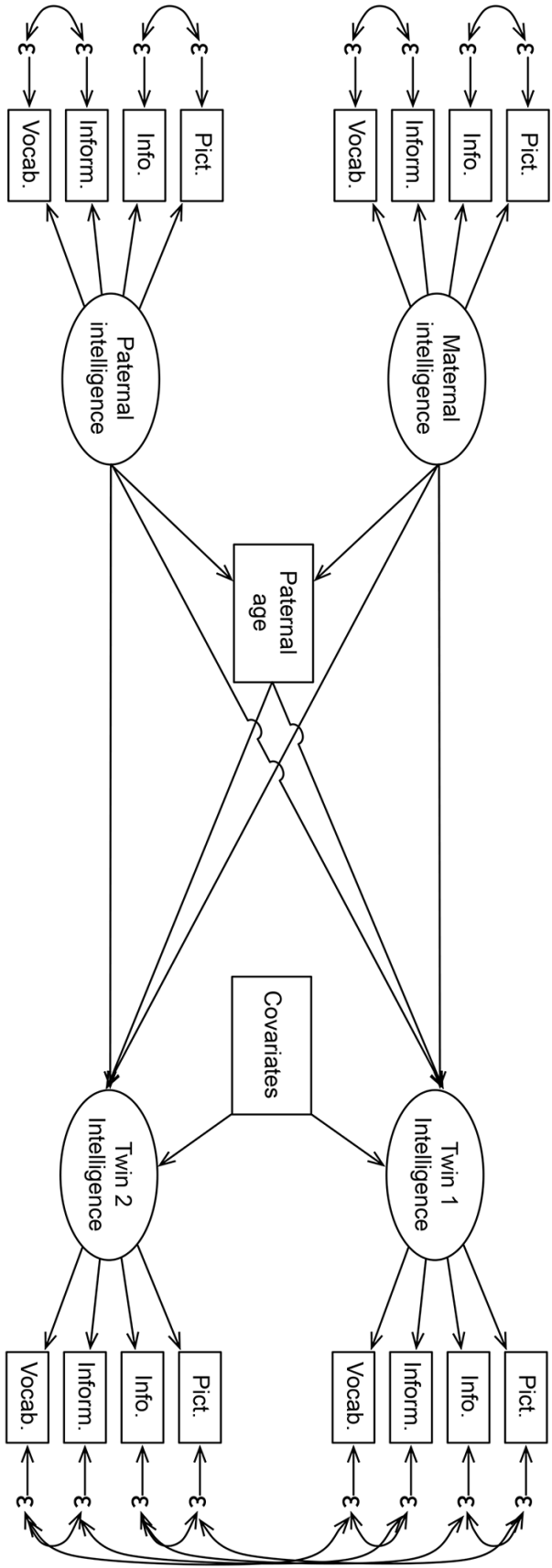

**Figure 1:** Simplified illustration of the structural equation model. Other models were largely analogous, exceptions are explained in the text.

In the first model, we included methodological controls, namely child's sex and age at testing, to decrease residual variance and increase predictive power, and zygosity to control for correlated prenatal factors. The main predictor was paternal age at birth in days. In the second model we added parental trait levels. From the second intelligence model on we also added parental years of education as auxiliary variables, to improve the FIML estimation of missing intelligence data for parents. To compare our methods with previous studies we estimated models controlling only for either parental education, or intelligence or both. In the third model we added the number of older non-co-twin siblings (i.e. birth order), birth weight and birth complications as further controls.

For all analyses we chose one twin from each pair at random and then tried to replicate the result with the co-twin data. We also modelled quadratic trends emulating Malaspina et al.'s [76] analyses, and cubic trends for paternal age as suggested by Crow [40].

Furthermore, we examined associations with the primary scales of MPQ personality using multiple regressions and verbal and performance intelligence using a SEM. We ran analyses with and without the enrichment sample.

Complete, reproducible reports of the analyses have been made available online at http://openscienceframework.org/project/wLrZF/.

## 3.  Results

Including the enrichment sample made some results reach the conventional level of statistical significance but did not change the pattern of results in a noteworthy manner, so we opted for including it to reach higher power. We did a power analysis with G*Power 3 [121] to compute our study's sensitivity at a power of 95% and a Type I error probability of 5% to estimate the upper-bound effect size that could be detected. Our sensitivity analysis using the $n$s of complete cases indicated that we would be able to find paternal age effects if they explained at least 0.85% of the variance of IQ, 1.06% of head circumference variance, and 1.30% of MPQ personality superfactor score variance. Sensitivity in the FIML analyses would be higher.

The average paternal age at twin birth was 30.15 years ($SD$ = 5.54, range = 15-53) and fathers were born between 1925 and 1977. Mothers were born about 2 years later on average and twins were born on average in 1982. Mothers reported birth complications for 51% of all twin births and an average birth weight of 2587 grams ($SD$ = 563). Twins averaged slightly less than one older and one younger sibling, though relatively rarely in the same family. Parents averaged about 2 years of post-high school education and IQs very slightly above average and twin IQs were similar. Descriptive statistics for the other main variables can be found in Table 1. Parent-offspring correlations for IQ ($r$s = .39) and head circumference ($r$s = .24-.28) were moderate but somewhat lower for MPQ personality ($r$s = .10-.21). Correlations between mothers' and fathers' traits were similar (IQ: $r$ = .34; MPQ: $r$s = .15-.21), except for head circumference, which was effectively zero ($r$ = .04).

    Model fit according to root-mean-square error of approximation (RMSEA; both for baseline and full model) and standardized root-mean-square residual (SMRR) are reported in Table 2. Model fit according to $\chi^2$ was always violated owing to the large sample. The measures of close fit for the intelligence models exceeded recommendations by Browne and Cudeck [122]. The MPQ models' fit can still be regarded as reasonable for a parsimonious model because we did not model cross-loadings that were not part of the theoretical factor structure.

**Table 2:** Fit indices for the reported models

| Model | $\chi^2$ | $df$ | RMSEA [90% CI] | SRMR |
|---|---|---|---|---|
| Intelligence | | | | |
| 1. | 210.81 | 47 | 0.04 [0.04, 0.05] | 0.03 |
| 2. | 828.78 | 199 | 0.04 [0.04, 0.04] | 0.04 |
| 3. | 978.91 | 271 | 0.04 [0.04, 0.04] | 0.04 |
| Personality | | | | |
| 1. | 3323.16 | 233 | 0.08 [0.08, 0.09] | 0.08 |
| 2. | 7312.95 | 960 | 0.06 [0.06, 0.06] | 0.06 |
| 3. | 7799.78 | 1130 | 0.06 [0.06, 0.06] | 0.06 |

Note. All reported $\chi^2$ were significant ($p < .001$). $N = 1898$

$df$ = Degrees of freedom

RMSEA = Root mean squared error of approximation

SRMR = Standardized root mean residual

CI = Confidence interval

Model 1: Paternal age, twin's age at testing, sex, zygosity.

Model 2: As model 1, plus mother's trait level, father's trait level.

Model 3: As model 2, plus number of older siblings/birth order, birth weight, birth complications.

**Table 3:** Standardized regression coefficients for paternal age in three models.

| Variable (each twin) | Model 1 $b$ | [95% CI] | Model 2 $b$ | [95% CI] | Model 3 $b$ | [95% CI] |
|---|---|---|---|---|---|---|
| Full IQ[a] | 0.12*** | [0.06, 0.17] | -0.04† | [-0.09, 0.01] | -0.01 | [-0.06, 0.04] |
| [b] | 0.11*** | [0.06, 0.16] | -0.04 | [-0.09, 0.01] | -0.02 | [-0.07, 0.03] |
| Head circumference[a] | 0.04 | [-0.01, 0.09] | 0.03 | [-0.02, 0.08] | 0.03 | [-0.02, 0.08] |
| [b] | 0.03 | [-0.03, 0.08] | 0.02 | [-0.03, 0.07] | 0.01 | [-0.03, 0.06] |
| Positive affectivity[a] | -0.03 | [-0.08, 0.03] | -0.03 | [-0.08, 0.03] | -0.02 | [-0.07, 0.04] |
| [b] | -0.03 | [-0.09, 0.03] | -0.03 | [-0.08, 0.03] | -0.02 | [-0.08, 0.04] |
| Negative affectivity[a] | 0.01 | [-0.05, 0.08] | 0.04 | [-0.02, 0.10] | 0.03 | [-0.04, 0.09] |
| [b] | 0.02 | [-0.04, 0.08] | 0.04 | [-0.02, 0.10] | 0.02 | [-0.04, 0.08] |
| Constraint[a] | 0.02 | [-0.05, 0.09] | -0.03 | [-0.09, 0.04] | -0.02 | [-0.09, 0.06] |
| [b] | 0.02 | [-0.05, 0.09] | -0.01 | [-0.08, 0.06] | 0.00 | [-0.08, 0.07] |
| Absorption[a] | 0.04 | [-0.01, 0.09] | 0.06* | [0.00, 0.11] | 0.06* | [0.00, 0.11] |
| [b] | 0.05† | [0.00, 0.10] | 0.06* | [0.01, 0.11] | 0.07* | [0.01, 0.12] |

Note. Latent variables were standardized. Coefficients are the change in outcome in standard deviation units per decade of paternal age. No adjustment of significance levels for multiple testing.

Model 1: Paternal age, twin's age at testing, sex, zygosity.

Model 2: As model 1, plus mother's trait level, father's trait level.

Model 3: As model 2, plus number of older siblings/birth order, birth weight, birth complications.

[a]: Twin 1; [b]: Twin 2

CI = Confidence interval. † $p < .10$. * $p < .05$. ** $p < .01$. *** $p < .001$

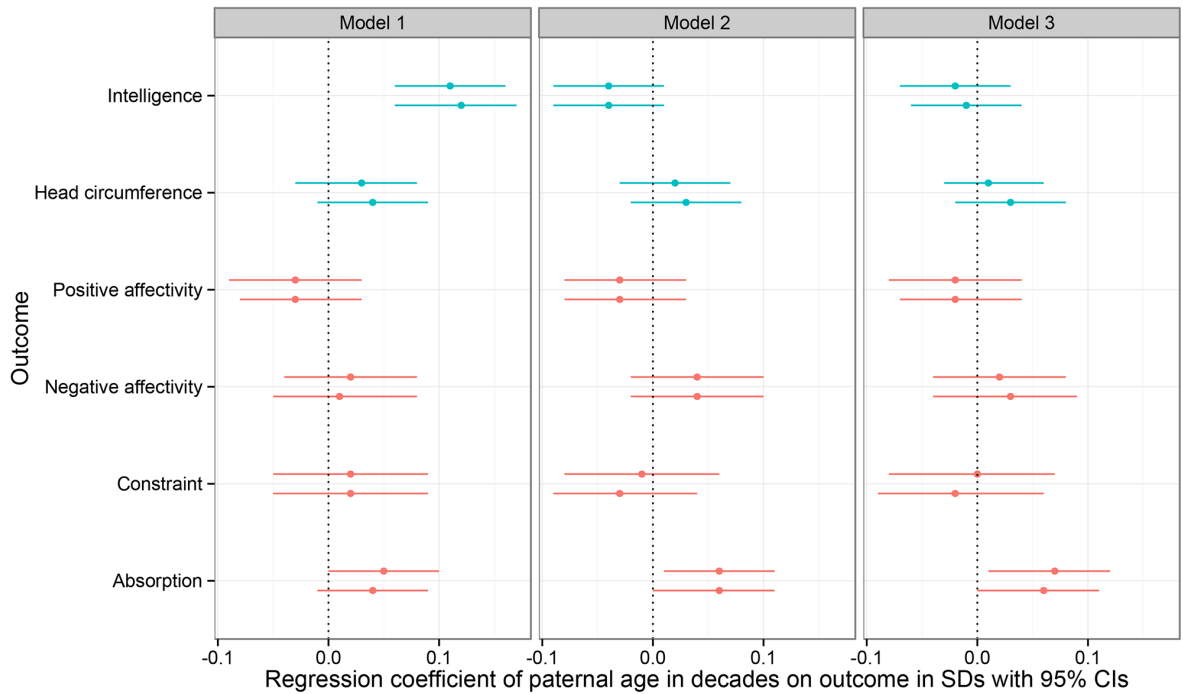

**Figure 2:** Regression coefficients of paternal age on main outcomes plotted for the two model chains. Same colours indicate coefficients estimated in joint models. Twins are presented separately, with the first co-twins presented first.

Overall, we found no robust evidence for paternal age associations with intelligence, personality or head circumference (see Table 3 and Figure 2). The regression coefficient of paternal age on offspring intelligence turned from positive to negative after controlling for parental intelligence, because both predictor and outcome correlated positively with parental intelligence. The change from b = 0.12 (95% CI [0.06, 0.17]) in the first model to b = -0.04 (95% CI [-0.09, 0.01]) was significant. The confidence intervals overlapped if we controlled for years of education instead of intelligence in the second model (b = 0.04, 95% CI [-0.01, 0.09], see also Figure 3). Controlling for both education and intelligence yielded a descriptively larger change than controlling for either one in the regression weight that reached conventional significance (b = -0.06, 95% CI [-0.11, 0.00]). In the third model, correcting for birth order decreased the relation to statistical non-significance (tested by omitting other covariates post-hoc). The regression weights for birth order were -0.16 (p < .001, 95% CI [-0.21, -0.11]) and -0.11 for the co-twins (p < .001, 95% CI [-0.16, -0.06]).

We also found a positive relation between paternal age and Absorption that was significant for both twins in the second and the third model, but not without the enrichment sample. We found no statistically significant relation of paternal age with the MPQ superfactors or with head circumference.

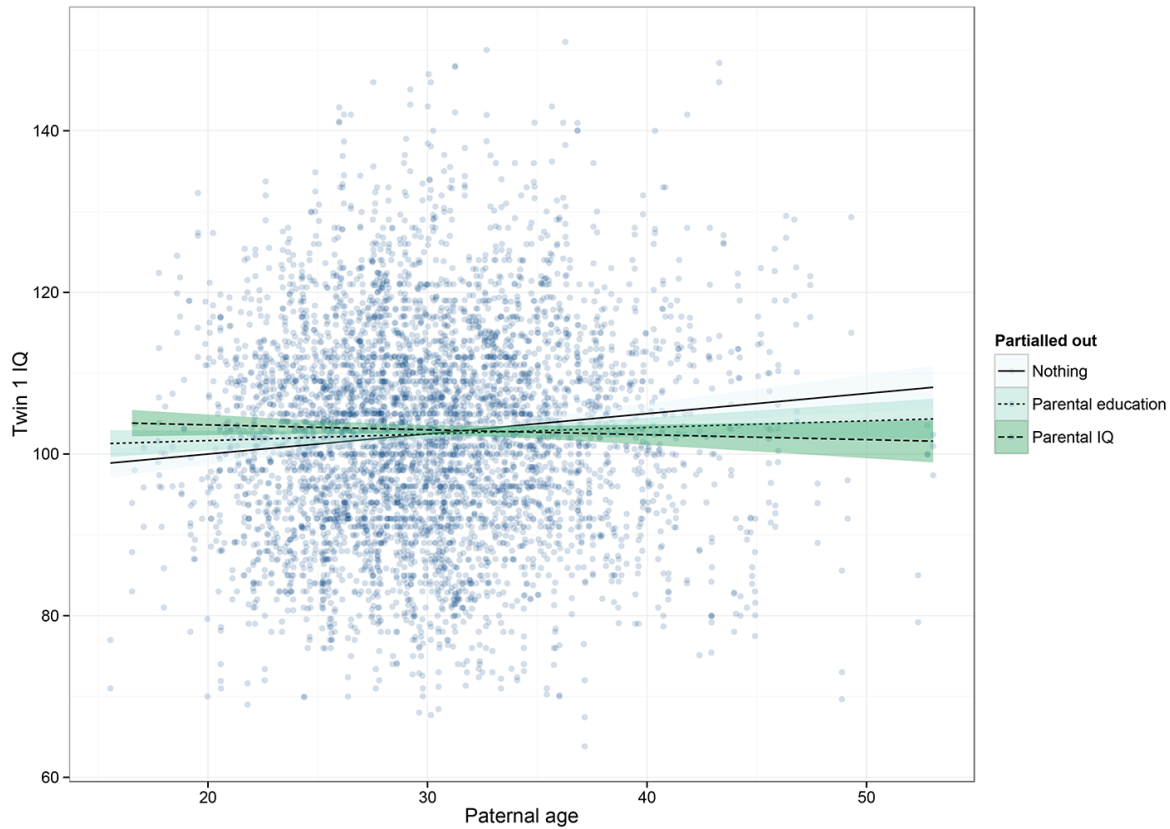

**Figure 3:** Raw data of the association between paternal age and offspring IQ (only complete cases). Superimposed are three fit lines, with different covariates partialled out of paternal age.

We tested for confounding relations between paternal age and parental traits. In multiple regressions estimated in the SEMs we found positive regression coefficients for maternal ($b = 0.17$, $p < .001$, 95% CI [0.11, 0.22]) and paternal intelligence ($b = 0.15$, $p < .001$, 95% CI [0.09, 0.20]) on paternal age, indicating later reproduction among more intelligent parents. MPQ personality was also related to paternal age. Paternal positive ($b = -0.10$, $p = .014$, 95% CI [-0.19, -0.02]) and negative ($b = -0.11$, $p = .012$, 95% CI [-0.20, -0.03]) affectivity were related to lower paternal age. Other non-significant, but not significantly less important predictors in this regression were maternal negative affectivity ($b = -0.05$, $p = .073$, 95% CI [-0.11, 0.00]) and paternal constraint ($b = 0.06$, $p = .123$, [-0.01, 0.13]). Parental intelligence explained significantly more variance in paternal age ($R^2 = 0.07$, 95% CI [0.05, 0.10]) than parental MPQ personality ($R^2 = 0.02$, 95% CI [0.00, 0.04]). Parental education also predicted paternal age (maternal $b = 0.15$, 95% CI [0.10, 0.21], $p < .001$; paternal $b = 0.06$, 95% CI [-0.00, 0.12], $p = .048$), though intelligence explained more variance on its own than education ($R^2 = 0.04$, 95% CI [0.02, 0.05]). Together they did not account for significantly more variance ($R^2 = 0.08$, 95% CI [0.05, 0.10]). In a joint regression on paternal age estimated as part of the second model, the coefficients for maternal intelligence, paternal intelligence, maternal and paternal education respectively were 0.15 ([0.07, 0.23], $p < .001$), 0.16 ([0.07, 0.26], $p = .001$), 0.06 ([-0.01, 0.12], $p = .125$) and -0.07 ([-0.15, 0.01], $p = .096$).

Examining quadratic and cubic effects of paternal age, as well as tests at the subtest and primary scale level, did not yield many noteworthy results. Modelling verbal and performance intelligence separately in the second model showed mostly overlapping 95% CIs of [-0.11, 0.00] for verbal and [-0.10, 0.03] for performance intelligence. In the third model we found a quadratic association (linear $b = -0.04$, p = .148, 95% CI [-0.09, 0.01]; quadratic $b = 0.07$, $p = .010$, 95% CI [0.02, 0.12]) of paternal age with offspring verbal IQ that replicated for the co-twins (linear $b = -0.05$, p = .055, 95% CI [-0.11, 0.00]; quadratic $b = 0.07$, $p = .008$, 95% CI [0.02, 0.13]), but it was in the opposite direction of what we had predicted. We tested our results' robustness to leaving out covariates and other modelling decisions such as using FIML instead of multiple imputation, or imposing measurement invariance according to Raykov et al. [123]. With the exception of the covariates birth order and parental traits in the intelligence models, this did not lead to noteworthy changes in the results pattern.

## 4. Discussion

We did not find support for our hypothesis that higher paternal age at offspring conception, as an indicator of more new, harmful mutations, would predict lower offspring intelligence. A small positive association between paternal age and offspring intelligence turned significantly negative after controlling for parental intelligence and education, but this finding was not robust to adding birth order as a covariate, leaving out the enrichment sample, or correction for multiple testing. We found small positive relations between parental intelligence and both paternal and maternal ages, plausibly indicating delayed reproduction among higher-IQ parents. Unlike Rodgers et al. [15] and Neiss et al. [124], who reported that education mediated the relation between maternal intelligence and female age at first birth, we found that parental education did not account for a significant amount of variance in paternal age over and above parental intelligence. This might indicate that paternal and maternal ages at twin birth were not representative of maternal age at first birth (the most commonly used indicator of reproductive timing). We think it is unlikely that this discrepancy reflects deeper underlying differences with regard to reproductive planning in our twin sample as twin births are not usually planned. Differential utilisation of assisted reproductive technologies (ART) would further complicate the picture, MTFS twins, however, were born at a time when ART were less common reasons for multiple births [125,126].

Unexpectedly, we found a significant association with one MPQ scale, Absorption, applying no adjustments to significance level for the multiple testing we did. To the extent this association might be real, we speculate that it might reflect the well-replicated association of paternal age with offspring schizophrenia, because Absorption has been found to correlate with clinically aberrant experience [127], hallucinations [128] and PSY-5 Psychoticism [129]. Although a potential link with new mutations, as indicated by the parental age association, could explain why Absorption has not been found to be elevated in mostly non-offspring kin of schizophrenia patients [130], we would only cautiously interpret this finding in the light of the fact that the association was not robust to correcting for multiple testing.

For the MPQ superfactors, constraint, positive and negative affectivity, we did not find any significant relations with paternal age, either before or after controlling for parental personality. Offspring head circumference was not significantly related to paternal age either. These results provide indirect support for our hypothesis that genetic variance in neither personality traits nor head circumference is under mutation-selection balance [8,9,109].

One of the primary strengths of this study was our ability to control for parental trait levels measured with the same precision as offspring traits when using paternal age as an indicator for likelihood of new mutations. Even though our sample was smaller than those of preceding studies, we would have been able to detect some of the previously reported effect sizes had they been present (e.g. Malaspina et al.'s 2% incremental variance explained [76]). However, some reported effect sizes would have been too small for us to detect and we cannot have too much confidence in power estimates derived from previous studies that probably suffered from varying degrees of omitted variable bias. Most importantly, neither the relation with intelligence nor the relations with constraint, positive or negative affectivity were significant. Lack of constraint coupled with negative affectivity is similar to externalising behaviour [131], which Saha et al. [93] reported to be related to paternal age. Positive and negative affectivity are related to social functioning [132], which Weiser et al. [92] found to be associated with paternal age. Despite the smaller size of our sample, we were able estimate the upper effect size boundary when controls for parental trait levels were in place, and can say with some confidence that true effects would not explain more than 1.3% of variance. Higher heritability does not imply that it will be easy to detect individual causal genes. However, with samples in which less variation is accounted for by non-genetic components (e.g. shared-environment), we would expect a paternal age effect attributable to mutations to explain more variation and thus to be more easily detected. This could for example be the case in samples with older offspring [133,134].

A large effect of paternal age on intelligence would have been consistent with a detrimental burden of new mutations coming from older fathers and would have thus raised the question why selection has not led to early reproduction (or even "andropause", i.e. a complete cessation of male reproductive ability in late life) in men. It would also have indicated selective pressure for transcription accuracy. Very small effects of paternal age are consistent [47] with the hypothesis that new mutations affecting fitness are rare and have small effects on the population level (though their effects on single individuals might still be substantial).

A link between paternal age and a trait in which variation is maintained through mutation-selection balance should persist or even emerge only after controlling for parental trait levels. Parents' intelligence and personality may influence both their reproductive timing and their children's traits, thus constituting an unobserved common cause of both paternal age and offspring traits. If we assume that the mean time at which the parents had the twins was representative of their mean overall reproductive timing (we were unable to test this beyond showing that parental intelligence was unrelated to twins' number of older or younger siblings), parents with higher IQs delayed reproduction compared with those with lower IQs in our sample. This is the most likely reason that the association between paternal age and offspring intelligence turned from positive to negative when controlling for parental intelligence, an effect that was not apparent when controlling only for parental education, as has been done in previous studies. Because controlling for education led to regression coefficients whose confidence intervals overlapped with those of the first and second models' and because our results regarding the importance of education to reproductive timing differed from previous studies' [15,124], we recommend controlling for both in samples in which fertility may be influenced by personality [4,135,136] and intelligence [15]. In particular, associations with reproductive timing have not yet been demonstrated in a sufficiently wide range of samples which differ with regard to family planning. At least in our sample we did not find any noteworthy changes in the regression coefficients of paternal age on personality when adding parental trait levels as covariates. Thus, it may be possible to assess effects of paternal age on personality in simple cross-sectional samples without having to account for the indirect path through the common cause parental personality.

An interaction between societal factors leading to delayed reproduction and IQ might explain that results differed in previous studies, though they used similar controls and methods. In the case of Auroux et al. [83,84] a largely overlapping research group working with French military recruit data found a negative effect of increasing paternal age on IQ, but could not replicate it in more recent data. If the societal trend towards delayed reproduction in industrialised countries [137] were accelerated in people with higher IQs, parental IQ, as an unobserved common cause in previous studies, would have suppressed the path from paternal age to offspring IQ more in studies of more recent lower-fertility cohorts. Saha et al. [78] and Malaspina et al. [76], who reported a negative association between paternal age and offspring IQ, analysed samples from populations with high average fertility. Average fertility rates in the USA and Israel were 3.6 and 3.8, roughly double those in France, Sweden, the United Kingdom and Minnesota (1.6 to 2.2; national fertility rates at the time of data collection from [138]; Minnesota fertility rates from [139]) in the respective birth cohorts of the studies that did not report negative associations [80,83,85]. Speculatively, any bias resulting from an effect of paternal IQ on both reproductive timing and offspring traits may have differed between these higher- and lower-fertility populations. Thus, societal fertility trends might account for differences among the studies in these different populations.

### Limitations

Our sample size was smaller than those of most previous studies (a fourth of Auroux et al.'s [83], a hundredth of Svensson et al.'s [80]), therefore our power to detect effects that explain less than 0.85% of variance was severely restricted. That our IQ tests were more established and comprehensive than the military aptitude tests and school grades used before can only partly compensate this. Because we analysed a rather small and homogeneous sample, our considerations regarding societal trends have to remain speculative and generalisability of results might be restricted. We also cannot know for sure whether paternal age at twin birth was representative of average reproductive timing and whether the associations we report would be replicated for single births. There is evidence against consequential mean differences in the outcomes of interest [104,140]. The relation between advanced maternal age and dizygotic twinning [141] would not, on its own, jeopardise our conclusions, though replication in a singleton sample would of course strengthen our confidence in them. The systematic differences we found for families whose fathers did not participate in the intake assessments may have decreased our chance to find significant results.

Our relative ability to detect any paternal age effects on MPQ personality as opposed to effects on intelligence may have been even lower than indicated by our sensitivity analyses, because we had less MPQ personality data, poorer model fit and less auxiliary information to estimate our models with missing data.

Because major disabilities and birth defects were thoroughly screened out of our sample, our conclusions are limited to intelligence variation in the normal range. Previous studies also conducted their analyses on either nonclinical or clinical samples, but not both. If paternal age were related to intellectual disability, but not intelligence in the normal range, effect sizes would also vary across studies according to the thoroughness of the screening procedure. The mean and variance of paternal age in our sample were similar to previous studies, but we cannot rule out that a larger number of older fathers would have boosted our explanatory power, especially if the effect were exponential.

We may also have omitted important confounding variables. Unlike previous researchers we decided against controlling for maternal age because this would have introduced high collinearity with paternal age and birth order. Explanations for offspring IQ increases with higher maternal age largely relied on child-rearing and parental psychological adjustment as mediators [142], for which we tried to account using parental IQs instead. Socioeconomic status was not controlled either, because we believed controls for intelligence and education to be sufficient.

**Considerations for future studies**

We hope future research on paternal age effects on intelligence will benefit from the debate about the effect of birth order on intelligence [102,103,143]. It seems possible to disentangle birth order and paternal age, because they generally have only moderate correlation across families. Some interpretations of the birth order variable (e.g. tutoring by siblings, or decreased paternal investment when multiple children are born in short intervals) would not be consistent with an effect of accumulated germline mutations, but e.g. decreased paternal investment in later-born siblings would be. Many of the challenges that emerged in birth order research apply to paternal age research as well. One example is the debate over whether birth order is also related to decreased intelligence within families [102]. If constant differences between families (e.g. parental intelligence) which are related to both their reproductive decision-making and mean offspring intelligence drive paternal age effects, they would be found between families, but not within. Such effects would not be indicative of new mutations and thus spurious in the context of our research question.

Of course within-family findings are not beyond reproach either [144]. For example, families may decide to have more children after their economic situations improve, allowing them to provide better environments for their later-borns. Additionally, within-family research may suffer from limited variance in paternal age, because most women do not have children across their whole reproductive lifespan in industrialised countries [145] and because fathers can only have children across their whole reproductive lifespans if they find younger partners after their original partner has gone into menopause.

Paternal age and birth order have the same rank-order within many "traditional" families. To break up this confound, the variable birth order could be substituted by direct assessments of the constructs for which it is supposed to be a proxy: differential parental investment, sibling tutoring and so forth.

Future research should also look into examining the paternal ages at births of previous generations (e.g. grandpaternal ages at parents' birth [146]) to obtain better estimates of recent additions to inherited mutation load. Pedigree analyses are the most realistic way to accomplish this, because grandparental trait information will rarely be available. Looking at the paternal ages of earlier generations would also enable researchers to tell apart the effects of mutations and germline epigenetic abberations, which might both underlie paternal age effects, because only epigenetic states can, potentially, be reset in grandchildren, therefore epigenetic effects would be expected to diminish more in grandparental analyses than expected based on relatedness [147].

Similarly, whole-genome- and exome-sequencing studies of families, which allow for counting new mutations by comparing the genomes or exomes of parents and children, have to control the various factors, especially parental trait levels, that might influence reproductive timing and thereby new mutation incidence. For example, Iossifov et al. [86] found that "likely gene-disrupting mutations" predicted autism, but were not related to intelligence in an exome-sequencing study of 343 families with children on the autism spectrum and their unaffected siblings. Inherited subsyndromal autism was an unlikely confound for the autism finding, because they employed a simplex sample (i.e. no relatives with autism spectrum disorders). However, parental intelligence was not controlled. The same concern applies to Sanders et al.'s [54,56] results, which implicated new copy-number but not single nucleotide variants in intelligence in a clinical sample from the same population, the Simons Simplex Collection. Most importantly, the exome constitutes only the coding 1% of the genome; plausibly more polymorphisms affecting complex, continuous traits may be found in the regulatory sequences of the genome [148]. The total contributions of new mutations to intelligence are only beginning to come into the reaches of current molecular genetic methods (especially the still expensive sequencing techniques, see [32,53,73,74]).

Another way paternal age studies can improve their estimates is by considering the insights from Flynn effect research (Flynn [149,150], reviewed by Mingroni [151]). The rise in intelligence test scores over time could mean that older parents in previous studies were also from earlier cohorts with lower test scores. Possibly, their offspring would have lower test scores as well. Malaspina et al. [76] dismissed the Flynn effect as a confound, because it had not been found to occur within families, nor to affect heritability estimates for intelligence. However, the Flynn effect has since been shown in brothers [152]. Johnson, Penke, and Spinath [153] reasoned that high heritability of a trait should not be construed as an argument against environmentally mediated secular increases: Gene-environment interactions may be revealed or hidden, depending on whether the necessary variability in the environment is present. Wicherts et al. [154] showed that measurement invariance of general intelligence was violated with respect to different cohorts, making it unlikely that the observed gains reflected latent "real" increases. Previous studies which used sum scores could not guard against bias resulting from changes in subtest scores rather than general intelligence by checking their results' robustness to imposing measurement invariance.

A paternal age effect could also mask a rise of test scores within families. Rodgers [155] had dismissed both the within-family Flynn and the birth order effect, arguing that neither was present in his data, even though the two effects might have cancelled each other out [151].

In fact, Sundet, Borren, and Tambs [152] have proposed changes in fertility patterns as one cause of the Flynn effect after finding that decreases in the prevalence of large families explain part of the increase in intelligence scores. They examined data on Norwegian conscripts, but they aggregated mean sibling IQ. Plausibly the actual explanatory variable is found elsewhere, at the individual level. A trend towards delayed reproduction in intelligent parents [15] and the general population [137], and thus an increase in new mutations, could be partly culpable for the reports of a slowing [156], stop [157] or even reversal [158,159] of the Flynn effect in Scandinavian countries. We might be able to explain the null effects of paternal age on intelligence in more recent analyses, in which parental intelligence was not controlled [80,83,85] by delayed reproduction among more intelligent parents, though our study raises the question whether any paternal age effect attributable to mutations exists and is substantial. Taking into account these known problems with measuring differences in intelligence over time could serve to improve future research into paternal age.

Additionally, research in more diverse populations is warranted because results from the 1000 Genomes Project Consortium [148] suggest that populations are substantially differentiated geographically with regard to low-frequency variants. The results also suggest differences in strength and efficacy of purifying selection across populations, which are highly relevant to paternal age research.

Future research on paternal age effects may benefit from the history of birth order research and employ controls for parental traits, within-family designs (eliminating between-family confounds) or pedigree analyses of paternal age effects across several generations (ruling out alternative environmental and epigenetic explanations and boosting explained variance) depending on the availability of data.

**Conclusions**

Controlling for parental trait level, we were unable to show significant effects of paternal age, a proxy for new genetic mutations, on offspring IQ, head circumference, or personality traits. Parents' IQ and personality were correlated with their reproductive timing. This necessitates thorough control of parental trait levels in future studies on paternal age effects. Our sample size was insufficient to reveal very small effects, but our results can be understood as providing an upper boundary of any expected effect sizes. Reported effect sizes of paternal age on offspring personality and intelligence have been heterogeneous. So far no clear picture of the role of mutation-selection balance has emerged from these studies. More research in different populations and converging evidence may enable us to find out more about the evolutionary mechanisms that maintain genetic variance in traits like intelligence. If any paternal age effects on intelligence exist, they are probably very small. Narrowing down the precise effect size and ruling out the many possible confounds would be steps towards quantifying the contribution of de novo mutation-selection balance to intelligence and other individual differences. If other studies show paternal age effects on intelligence to be negligible but confirm the link between paternal age and de novo mutations, this prompts interesting research questions into the robustness of the highly polygenic intelligence trait.

## 5. Acknowledgements

The first author wishes to thank Roos Hutteman and Sarah J. Lennartz for providing helpful feedback on an earlier version of this manuscript.